\documentclass[conference]{IEEEtran}

\IEEEoverridecommandlockouts
\usepackage{cite}
\usepackage{amsmath,amssymb,amsfonts}
\usepackage{algorithmic}
\usepackage{graphicx}
\usepackage{textcomp}
\usepackage{xcolor}
\def\BibTeX{{\rm B\kern-.05em{\sc i\kern-.025em b}\kern-.08em
    T\kern-.1667em\lower.7ex\hbox{E}\kern-.125emX}}

\makeatletter
\newcommand{\linebreakand}{%
  \end{@IEEEauthorhalign}
  \hfill\mbox{}\par
  \mbox{}\hfill\begin{@IEEEauthorhalign}
}
\makeatother

\begin{document}

\title{Definition and clinical validation of Pain Patient States from high-dimensional mobile data: application to a chronic pain cohort\\
}

\author{\IEEEauthorblockN{1\textsuperscript{st} Jenna M. Reinen}
\IEEEauthorblockA{\textit{Digital Health} \\
\textit{IBM Research}\\
Yorktown Heights, NY \\
jenna.reinen@ibm.com}
\and
\IEEEauthorblockN{2\textsuperscript{nd} Carla Agurto}
\IEEEauthorblockA{\textit{Digital Health} \\
\textit{IBM Research}\\
Yorktown Heights, NY \\
carla.agurto@ibm.com}
\and
\IEEEauthorblockN{3\textsuperscript{rd} Guillermo Cecchi}
\IEEEauthorblockA{\textit{Digital Health} \\
\textit{IBM Research}\\
Yorktown Heights, NY \\
gcecchi@us.ibm.com}
\and
\linebreakand
\IEEEauthorblockN{4\textsuperscript{th} Jeffrey L. Rogers}
\IEEEauthorblockA{\textit{Digital Health} \\
\textit{IBM Research}\\
Yorktown Heights, NY \\
jeffrogers@us.ibm.com}
\and
\IEEEauthorblockN{5\textsuperscript{th} NAVITAS and ENVISION Studies Physician Author Group}
\IEEEauthorblockA{\textit{Clinical Research} \\
\textit{Boston Scientific}\\
Valencia, CA \\
}
\and
\linebreakand
\IEEEauthorblockN{6\textsuperscript{th} Boston Scientific Research Scientists Consortium}
\IEEEauthorblockA{\textit{Data Research and Engineering} \\
\textit{Boston Scientific}\\
Valencia, CA \\
}
}
\makeatletter
\def\ps@IEEEtitlepagestyle{
  \def\@oddfoot{\mycopyrightnotice}
  \def\@evenfoot{}
}
\def\mycopyrightnotice{
  {\footnotesize
  \begin{minipage}{\textwidth}
  Copyright~\copyright~ 2022 IEEE. Personal use of this material is permitted. Permission from IEEE must be obtained for all other uses, in any current or future media, including reprinting/republishing this material for advertising or promotional purposes, creating new collective works, for resale or redistribution to servers or lists, or reuse of any copyrighted component of this work in other works.
  \end{minipage}
  }
}
\maketitle

\begin{abstract}
The technical capacity to monitor patients with a mobile device has drastically expanded, but data produced from this approach are often difficult to interpret. We present a solution to produce a meaningful representation of patient status from large, complex data streams, leveraging both a data-driven approach, and use clinical knowledge to validate results. Data were collected from a clinical trial enrolling chronic pain patients, and included questionnaires, voice recordings, actigraphy, and standard health assessments. The data were reduced using a clustering analysis. In an initial exploratory analysis with only questionnaire data, we found up to 3 stable cluster solutions that grouped symptoms on a positive to negative spectrum. Objective features (actigraphy, speech) expanded the cluster solution granularity. Using a 5 state solution with questionnaire and actigraphy data, we found significant correlations between cluster properties and assessments of disability and quality-of-life. The correlation coefficient values showed an ordinal distinction, confirming the cluster ranking on a negative to positive spectrum. This suggests we captured novel, distinct Pain Patient States with this approach, even when multiple clusters were equated on pain magnitude. Relative to using complex time courses of many variables, Pain Patient States holds promise as an interpretable, useful, and actionable metric for a clinician or caregiver to simplify and provide timely delivery of care.
\end{abstract}


\section{Introduction}

Recent advances in digital medicine have provided the opportunity to collect large sets of clinical data to evaluate and predict critical medical outcomes. For instance, mobile-based applications, accelerometers, and biosensors are now ubiquitous in phones and watches, enabling one to longitudinally track variables like mobility and speech, and facilitate patient symptom self-report. Importantly, these features may associate with clinical meaning. Large-scale studies have shown that data from mobile applications tracking daily activity may predict outcomes relevant to health and illness, such as in geriatric care and diabetes \cite{Ekelund2009}, \cite{Smirnova2020}. Further, language can assess affective, psycholinguistic, physiological, and cognitive features can predict physiological and pharmacological \cite{Agurto2020}, psychiatric \cite{Corcoran2018a}, and cognitive disease states \cite{Eyigoz2020}. These types of findings have demonstrated the promise of digital health profiles in understanding patient experience and predicting important clinical outcomes.

Despite these advances, the size and complexity of the clinical data generated by mobile applications is nontrivial to interpret and apply for several reasons. First, digital healthcare data can exist in multiple formats, creating the need to fuse vast amount of diverse information \cite{Yu2019}. Second, there is a need for methods that can obtain clear data representations. These methods should provide interpretation that are manageable in size, yet can maintain the characteristics of the raw information, allowing for patients and healthcare professionals to interpret and use the output \cite{Ambigavathi2018}. Attempts to reduce and understand such data in a biological context have commonly used data-driven methods, especially those using machine learning algorithms. This approach offers the advantage of being able to handle large, multidimensional data sets through the ability to recognize patterns or joint representations that are otherwise difficult to identify using standard statistical approaches, providing knowledge discovery about a particular topic that can span time, location, and scales \cite{Dinov2016}. In particular, clustering analysis offers the ability to collapse across otherwise incomprehensible multidimensional data and observe how features co-occur. In the case of spectrum illnesses that incorporate a range and variety of symptoms, decomposition can be helpful, with some outputs having an advantage in outcomes prediction \cite{Reinen2018}. But not all results allow for interpretation, and it is particularly susceptible to problems in small, unvalidated datasets which may result in overfitting and thus results that are not replicable or generalizable. Further, while results from unsupervised approaches may reveal meaningful clinical patterns, few methods exist to formally assign labels, rank, or identify qualitative aspects from the results of data-driven approaches through independent validation. 

A prime illustration of this problem is in chronic pain, a disease affecting a substantial percentage of the population \cite{Zelaya2019} that significantly impacts general function including employment, mental health, and social interaction. This heterogeneous condition interacts with well-characterized facets of health, including mood, sleep, psychosocial function, medication use, and mobility. However, the current practice for most pain studies is to evaluate outcomes based on pain magnitude alone, which does not consider all of the variance shown to predict treatment success, quality-of-life (QoL), or other measurements of physical, psychological, and social well-being \cite{Gatchel2007}. But, using all of these features as outcome variables is nontrivial to compute, conceptualize, and interpret. Few standard approaches have been developed that incorporate both the computational methodology required to complete such a task, and the ability to provide a clinically interpretable summary of the output. To date, machine learning has been used to predict pain outcomes, identify clinical subgroups \cite{Mullin2021}, extract knowledge, and detect structure in biological and clinical features \cite{Lotsch2018}. In chronic pain, while artificial intelligence (AI) has been applied to improve diagnoses, fewer studies apply it to the treatment and management of pain patients \cite{Jenssen2021}, and analyses that use longitudinal data or clinical validation are extremely limited. 

Given the quickly expanding capacity of digital health and learning algorithms to inform treatment outcomes in complex illnesses, there is a benefit to developing an approach to validate health states from multidimensional data. While it is known that various chronic pain symptoms can co-occur, it remains currently unknown whether symptom profiles may be successfully organized into distinct health states. Here, we propose a method by which we aim to identify clusters from high-dimensional, longitudinal data in chronic pain patients, and label them as Pain Patient States that may be operationalized for clinical application and decision making \cite{Reinen2020}\cite{Anitescu2022}. To this end, we examined data from chronic pain patients in three subsets of data: 1) with questionnaires only; 2) with questionnaires plus voice data; and 3) with questionnaires plus actigraphy data. The dimensionality of each dataset was reduced into stable clusters using standard unsupervised clustering algorithms. Next, we quantitatively evaluated the clusters based on relationships to established health metrics, using standard assessments as clinical benchmarks in chronic pain to compare the data-driven results. A clear ordinal rank of states emerged, allowing us to assign unique qualitative labels even in clusters that were nearly identical in pain magnitude, so that they may be used as clinically-informed states. This system serves as an example of organizing diverse types of large datasets and anchoring them to known metrics as to evaluate treatment or assess function. Here, these formerly convoluted data patterns may now act to contextualize signal, rank results, track longitudinal health changes, and monitor meaningful medical outcomes.

\section{Methods}

\subsection{Participants and Data Collection}

Participants were recruited from pain clinics in on-going, longitudinal, multi-center, clinical studies (Clinicaltrials.gov ID: NCT01719055) aimed to understand chronic lower back and leg pain patients who are candidates for spinal cord stimulator (SCS) treatment (Boston Scientific, Valencia, CA). Participants were recruited and enrolled in the NAVITAS and/or ENVISION studies at multiple United States clinical sites if they intended to receive or had already received an SCS trial or implant, were at least 18 years old, and had been diagnosed with intractable chronic neuropathic pain. Additionally, subjects may have been previously enrolled in the RELIEF study (Clinicaltrials.gov ID: NCT01719055). Data were included in this analysis from each study a subject was enrolled in. Health-related questionnaires were administered via an at-home, custom-designed clinical study version of a digital health ecosystem (Boston Scientific, Valencia, CA) for up to 36 months. The questions chosen included pain-related subjective ratings, symptoms hypothesized to contribute to variability in pain ratings, as well as symptoms hypothesized to be impacted by pain, specifically pain magnitude, mood, sleep, alertness, medication use, and activity. Following enrollment, data were collected in separate in-clinic and at-home data streams. Mobile data analyzed here included voice recordings, as well as daily, self-reported symptom monitoring, with the option to respond more frequently if participants wished. In addition, subjects were asked to wear a smartwatch to assess mobility using accelerometer data (Galaxy Watch S2, Samsung USA, Menlo Park, CA with custom watch application, Boston Scientific, Valencia, CA). In-clinic assessments were collected at the baseline (enrollment) visit, and at 1-month, 3-month, 12-month, and optionally 24-month and 36-month visits following enrollment. In the present analysis, we used in-clinic assessments to evaluate QoL \cite{Group1990} and disability measured by the Oswestry Disability Index, or ODI \cite{Fairbank2000} questionnaires. 

\subsection{Voice data processing}

Voice recordings were collected from weekly recordings based on prompts aimed to understand the participant's experience with pain. Speech features for psycholinguistic, sentiment \cite{Pennebaker2015}, and acoustic characteristics \cite{Eyben2010}, \cite{DeJong2009} were extracted from the audio files using in-house and standard code. Age and sex were regressed from all features. Next, to reduce dimensionality of these features, a principal components analysis (PCA) was used ($var \geq 2\%$) to identify the decomposed components. These components were later included in a clustering analysis alongside the 6 features derived from the questionnaires.  

\subsection{Actigraphy data processing}

Effective mobility was derived from the watch-based actigraphy data. It is a novel metric of physical function and activity meant to reflect the duration and type of activity a person experiences beyond steps or activities of daily life. Rates of activity were calculated into categories for each participant throughout the day. These categories ranged from Zone 0 (e.g., resting, using a mobile device while seated) to Zone 4 (e.g., intense or repetitive motion or vigorous exercise) and were used along with the questionnaire data in the clustering analysis. 

\subsection{Data and Clustering Analysis}

For each participant, all available data was downloaded and selected based on days for which all subjective features from questionnaires (e.g., overall/leg/back pain, mood, sleep hours, sleep quality alertness, medication use for opioid/over-the- counter/non-opioid pain medication, activity interference due to pain, and activities of daily life), as well as actigraphy and voice data (where applicable) were present. Patients were included in the analysis regardless of time point in the study (e.g., baseline/enrollment, SCS trial period, follow-up), in the interest of observing a spectrum of pain-related variability and experience. However, the criteria for removing samples from the analysis consisted of: 1) any day missing a single data point, 2) any individual having fewer than 10 total complete data points, and when applicable 3) individuals who wore the smartwatch for less than 10 days. All question value responses were normalized prior to cluster analysis to equate the different subjective feature values across the individual question, and data distributions were inspected for abnormalities. Next, each question categorized to assess pain, sleep, and medication use were averaged to produce single composite scores for each modality; for activity, a difference score was taken between the two questions, in which we include a penalty that account with pain interferes with any overall activity. If any participant had answered more than one question on a certain day, the average of those responses was used to represent the daily value for that category. Participants were assessed for their average responses over time in order to determine the extent to which some participants responded more frequently than others, and the analysis was rerun without outliers to further ensure cluster stability.

Cluster definitions were calculated using a k-means clustering algorithm with Euclidean distance exploring up to cluster solutions for k = 10. Optimal k was determined using multiple methods including sum of squares distances and silhouette values, agglomerative analysis, and consensus clustering. To ensure clusters were similar across subsamples of participants exhibiting variability in number of responses included in the analysis, we repeated the analyses in varying samples of participants in which highly contributing participants (those with higher daily average responses) were excluded. Next, we employed an analogous approach to examine cluster solutions over the course of time. Generally, we expected the clusters to remain similar over time with some slight changes (e.g., higher pain prior to therapy) that would be evident in the cluster. With this in mind, cluster solution results were then visually inspected in order to ensure similarities in qualitative characteristics and are discussed in the results section.

\begin{figure}[htbp]
\centerline{\includegraphics{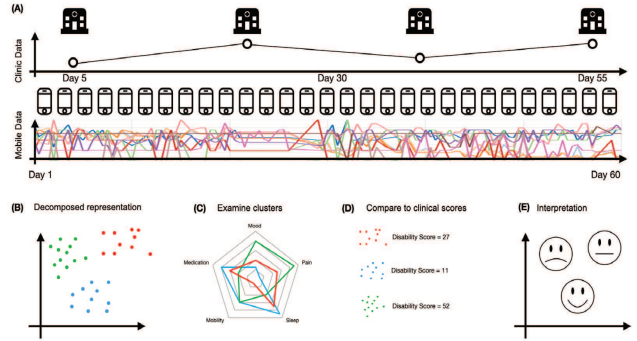}}
\caption{Conceptual data and methods overview. (A) Data were collected from a multi-center clinical trial recruiting participant with chronic low back and leg pain seeking spinal cord stimulator (SCS) treatment. Both in-clinic and at-home data collection were used to record 1) questionnaire-based daily reports of pain, mood, activity, medication, alertness, sleep; 2) standard assessments of QoL (EQ5D) and disability (ODI); 3) voice responses to open-ended questions about their pain; and 4) actigraphy from a smartwatch. (B) Data from questionnaires, voice, and actigraphy were subjected to a k-means clustering analysis and the (C) resulting cluster representation was examined across features. To validate these clusters, (D) centroid distance to each cluster was compared to the clinical scores for disability and QoL allowing for (E) an interpretation and label to be assigned to each cluster. }
\label{fig}
\end{figure}

\section{Results}

\subsection{Sample demographics and data chronology} 

In the primary analysis including questionnaires only, 121 individuals with 11,763 samples of data were used (40.5\% male, mean age 59.4 years old, 17.6 years since pain onset). In the analysis examining the addition of actigraphy data to the questionnaire data, 116 individuals with 11,286 samples of data were used (39.7\% male, mean age 59.3 years old, 17.8 years since pain onset). For the analysis including voice, 2,080 samples were included. 

\subsection{Clustering results and characteristics for questionnaire data only}

Cluster definition was examined for the questionnaire-only data for k = 2 to 10. Sum of squares distances and silhouette analyses indicated that a cluster solution of k = 2 or 3 was stable. Agglomerative hierarchical clustering was repeated to validate k with cross-methodological clustering, which also converged on a solution of k = 2 or 3. Given the relative stability of smaller cluster solutions, we first examined a simple and stable solution of k = 2. Feature characteristics of the cluster solution for k = 2 were examined by inspecting mean values for each feature in each cluster (Figure 2A). Results indicated a clear negative-to-positive grouping of health features, such that the questionnaire responses of one cluster appeared to represent a superior health state represented by better mood, sleep, alertness and activity, and lower ratings of pain and medication use. The other cluster appeared to represent an inferior health state, characterized by higher pain and medication use, with lower ratings for alertness, mood, sleep, and activity. This analysis was repeated to exclude the  high-responder group in order to ensure that the clusters were not being driven by the high-responders. Results indicated that the clusters were very similar both in all participants, and without the high-responders. Finally, the cluster solutions were re-examined over the course of time, such that the analysis was repeated in the baseline period prior to SCS activation, during the first 6 months of treatment, and the subsequent 6 months of treatment. Results indicated that the cluster solution was very similar over time, with some indication of higher pain prior to treatment. An examination of a 3-cluster solution revealed a third, intermediate cluster that represented a health state similar to or in between the two states represented in the two-cluster solution (Figure 2B). This cluster showed relatively high ratings of alertness, mood, and sleep, but with intermediate values for pain, activity and medication use. A repeated analysis excluding high-responders also showed an intermediate cluster, with values for each feature with a magnitude between the previous two clusters.

\begin{figure}[htbp]
\centerline{\includegraphics{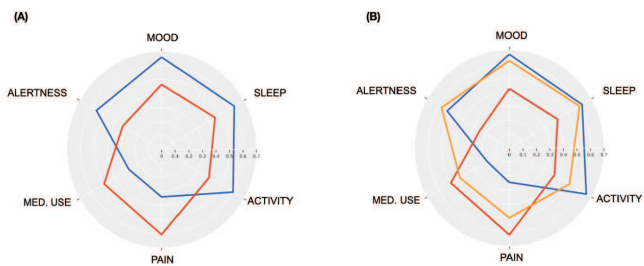}}
\caption{Cluster analysis for questionnaire data reveals negative and positive symptom groups. (A) A two cluster (k = 2) solution resulting from the k-means analysis of the questionnaire data revealed two clusters of symptoms that stratified on a negative-to-positive spectrum of pain-related health, in which one cluster revealed a better health state of better mood, sleep, more reported activity and alertness, less medication usage, and lower pain. Conversely, the other cluster depicted a worse health based on the feature means. (B) A three cluster (k=3) solution also revealed a spectrum of positive to negative symptom groupings, including superior and inferior states similar to k=2, with an additional intermediate state showing moderate pain and medication use but with high mood, sleep, activity, and alertness scores.}
\label{fig}
\end{figure}

\subsection{Decomposing and clustering questionnaire and voice data} 

Prior to clustering, the results of the principal components analysis (PCA) of the voice features were inspected. The results showed that 7 components were present, and characterized features such as voiced and unvoiced energy in a speech signal, negative sentiment, emotional content, and acoustic voice properties (Table 1). We repeated the clustering analysis with each of these 7 components included along with the 6 questionnaire components. With the addition of the 7 components, solutions for k of 2 or 5 were possible. For the k = 2 solution, results showed that in particular, component 4, which was characterized by high loadings of negative sentiment and acoustic features associated with emotion, tracked well with the inferior health cluster (Figure 3A). Further, while not all components showed the same discrimination between states as did component 4, there was evidence that the addition of the voice data expanded the granularity of the state solutions. This was illustrated by the comparison of a cluster solution with only questionnaires in which pain was stratified across 3 levels in all states (Figure 3B). When the cluster solution included both voice and questionnaires, pain across states expanded to 5 levels (Figure 3C). 

\begin{table}[htbp]
\caption{Decomposition of voice features into 7 components}
\centerline{\includegraphics{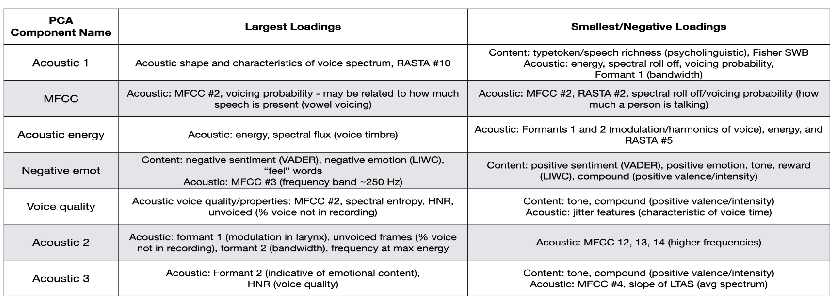}}
\label{fig}
\end{table}

\begin{figure}[htbp]
\centerline{\includegraphics{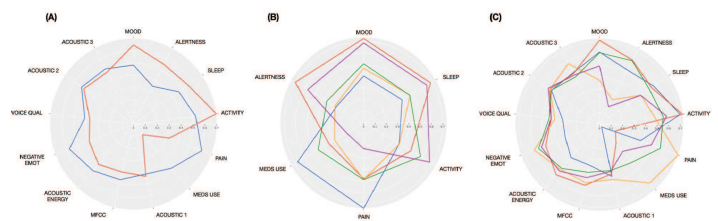}}
\caption{Adding voice features to cluster analysis improves pain granularity in state solutions. (A) Clustering analysis was run including 6 questionnaire components and 7 voice components for a 2-state solution, indicating that voice features denoting negative sentiment were associated with the poorer health cluster. (B) A 5-state cluster solution without voice features reveals three levels of pain magnitude across clusters, while the (C) addition of voice to a 5-state cluster solution adds further granularity to pain magnitude across clusters.}
\label{fig}
\end{figure}

\subsection{Clustering results and characteristics for questionnaire and actigraphy data}
Actigraphy data downloaded from the watch were parsed into mobility Zones 0 - 4 of effective mobility. Inspection of results indicated that these zones indeed provided granularity that added description beyond number of steps or self-reported ADLs (Figure 4). The clustering analysis included the 6 categories derived from the questionnaires along with the effective mobility. An analysis for optimal k showed that state solutions of up to 5 clusters was possible. These clusters appeared to range from a "best" state that included low pain and medication use, and high reports of mood, sleep, alertness, and effective mobility, to an inferior state that is associated with high levels of pain and medication use, and low reports of activity, mood, sleep, alertness, and effective mobility (Figure 5).  

\begin{figure}[htbp]
\centerline{\includegraphics{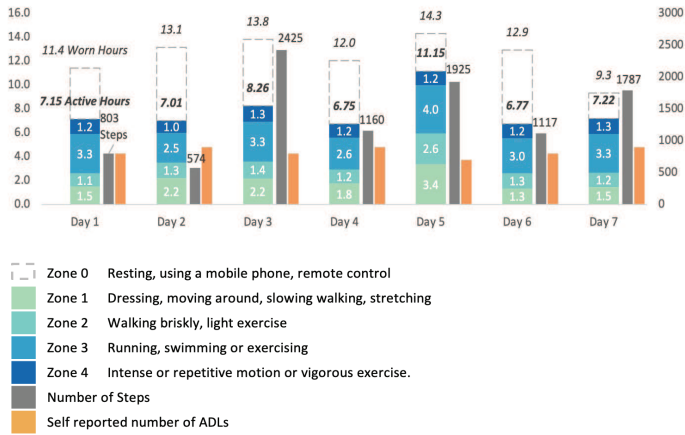}}
\caption{Description of effective mobility zones. Mobility data was parsed into zones of “effective mobility” based on rates of activity calculated at regular time window intervals throughout the day. When compared to step counts and self-reported activities of daily life (ADLs), effective mobility showed additional computational granularity of participant mobility.}
\label{fig}
\end{figure}

\begin{figure}[htbp]
\centerline{\includegraphics{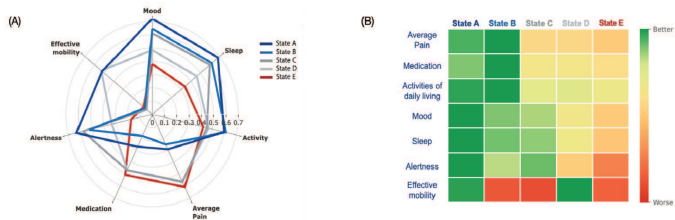}}
\caption{Adding mobility features contributes to cluster dimensionality. (A) A cluster solution including effective mobility identified 5 stable clusters for which the addition of effective mobility may contribute to additional clusters relative to the questionnaire-only solutions, still ranging from a negative-to-positive spectrum and including a best and a worst state. (B) States from the 5-cluster solution show further granularity as it pertains to patient experience beyond the 2- and 3-state model.}
\label{fig}
\end{figure}

\subsection{Cluster validation and state classification} 
For the validation analysis, we obtained pairs of metrics comprised of 1) distances from the cluster centroids on a given day; and 2) responses to standard assessments (disability, or ODI, and QoL, or EQ5D measurements focusing on Pain, Activities, and VAS Health). These two metrics were collected within one week of each other; any pairs with collection dates outside of the week window were dropped from analysis. We first calculated correlations between centroid distances of each cluster in the two-state solution, and found that the  correlations were statistically significant and consistent in terms of direction and magnitude for the two states, indicating a clear best and inferior state (in cluster 1 values were: disability/ODI, r = 0.42, EQ5D Pain, r = 0.47, EQ5D Activities r = 0.37, EQ5D VAS Health r = $-0.32$; all p-values $<$0.001, for cluster 2 values were: disability/ODI, r = $-0.41$, EQ5D Pain, r = $-0.43$, EQ5D Activities r = $-0.38$, EQ5D VAS Health r = 0.28; all p-values $<$ 0.001). This indicated that larger centroid distances were associated with higher values for the outcomes. Critically, while most of the validation metric outcomes represented negative health values with increasing severity including disability, EQ5D-Pain, EQ5D-Activities, etc., the EQ5D measure of VAS Health represents health on a positive scale, and as expected showed an inverse relationship to the findings above. Given that each cluster was associated with consistent directionality across all of the standard assessments, we were able to infer that each of the clusters represented distinct health states, aligned with what we would have expected to find in patients across time. 

\subsection{Cluster validation with voice data} 

A similar analysis was repeated using the k = 2 cluster solution that included voice data. Results indicated that generally the directionality of the correlations was consistent relative to prior analyses. However, for several validation metrics, the magnitude of the r values increased with the addition of voice features (for disability/ODI, r = 0.47, EQ5D VAS Health r = $-0.49$). In particular, assessments that may take into account negative affect showed an increase in the correlation across these metrics. Notably, because voice data is collected less frequently, there was a decrease in sample size relative to the prior analysis. That said, permutation tests were used to compare across the two approaches and to ensure that there were no meaningful differences due to sample size. In all instances, permutation tests confirmed the significance of prior findings at p $<$ 0.05. 

\subsection{Cluster validation with actigraphy data} 
Next, we aimed to determine whether correlations between centroids from a more highly dimensional state solution compared to the standard assessments could provide further ordinal information about the states. To do this, we ran a similar analysis using the 5-state solution that was obtained with the cluster solution including effective mobility. Here, we found that the correlations across the 5 states also provided evidence for a consistent ranking of those states from best to worst (Table 2). 

\begin{table}[htbp]
\caption{Cluster characteristics including effective mobility}
\centerline{\includegraphics{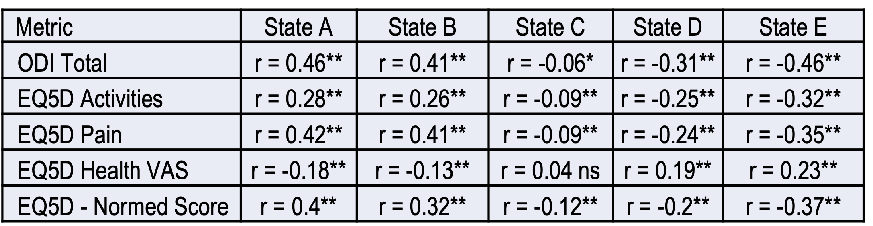}}
\label{fig}
\end{table}

\subsection{Comparison of state timecourse to health events}
In an exploratory analysis, we examined the relationship between state expression change over time relative to known health events. Here (see Figure 6), we first show that states represent a more interpretable visualization of health changes across time relative to examining the timecourse of all variables at once. Second, several exemplar patients show expected changes in states before and after implantation of the SCS device, a procedure that involves surgery and probable eventual pain relief. 

\begin{figure}[htbp]
\centerline{\includegraphics{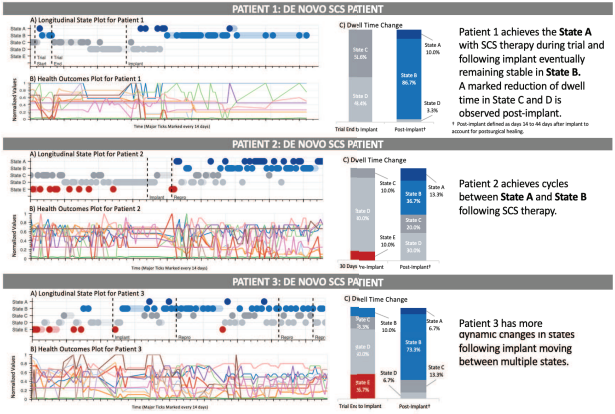}}
\caption{Examples of patient experiences show that states track with meaningful clinical events. Top time course for each patient denotes state assignment, whereas lower time course shows changes in multiple variables. Bar graphs show the dwell time change before and after a notable event, which here involves the implantation of a SCS device hypothesized to bring about eventual pain relief and improvement in QoL. (Here, data in the time courses included overall, leg, and back pain, sleep hours and quality, number of activities, pain interference, medication usage for opioid, over-the-counter, and non-opioid pain medications, alertness, mood, and effective mobility. States are ranked as $A>B>C>D>E$, as shown in Table 2.)}
\label{fig}
\end{figure}

\section{Discussion}

\subsection{High-dimensional health data can be decomposed meaningfully} 

Using a unique set of longitudinal questionnaire, mobility, and speech data, we have developed a novel method to decompose, group, and validate large amounts of chronic pain digital health data. This study marks one of the only approaches to create clinically usable pain-related categories from complex questionnaire, mobility, and speech data across time. This approach demonstrates that high dimensional, longitudinal health data from chronic pain patients may be decomposed into clusters and used to classify patients according to a holistic status named Pain Patient States. These states have an ordinal ranking based on clinically-validated standard health assessments. Specifically, we demonstrated that in chronic pain, we can take multiple streams of information including sleep hours and quality, mood, pain magnitude at multiple sites, alertness, multiple types of medication use, ADLs, actigraphy, and speech in order to represent 3-5 Pain Patient States over the course of time. The stable solutions that emerged from this method suggest the discovery of distinct clinical states with non-obvious properties that may serve as new knowledge that informs biological mechanisms and clinical care. In addition to the identification of these Patient Pain States, this improves upon prior assessments and clinical trials that only use pain magnitude as an outcome evaluation by considering a much more comprehensive picture of patient experience in a way that is clinically interpretable. This approach leverages both data- and clinically-driven analyses by first using powerful learning algorithms, and then comparing the output to standard clinical metrics. Consequently, we are able to transform what was previously multiple, complex time courses for hundreds of patients into 3-5 states that are clinically contextualized, straightforward, and meaningful. 

\subsection{The decomposition can be externally validated and ranked} 
We found that the resulting clusters from our analysis stratified on a negative-to-positive spectrum of health in chronic pain, and that these clusters were reliable across subsets of individuals and over time. Importantly, these states provide valuable, novel information per se, representing new findings that may define patient experience. Nevertheless, because they were derived from a purely data-driven analysis, we chose to compare cluster characteristics to independent standard assessments of disability and QoL. We found not only that good and bad clusters associate with better and worse disability and QoL, but that more granular state solutions had a clear ordinal rank which contextualized the data-driven output (Table 2). Further, in a 5-state solution (see figure 5A), only 2 levels of discriminable pain were present for 4 states. This adds clear dimensionality beyond what pain alone may indicate about a patient's well-being. Thus, we were able to assess 5 ordinal steps of health based on multidimensional aspects, providing evidence that we can offer a more full picture of patient experience yet preserve interpretability, making these states meaningful and actionable clinical information. This can improve precision in outcomes assessments, especially as it pertains to pain research and clinical trials. 

\subsection{Objective data adds granularity to state solutions} 
In particular, raw objective metrics such as actigraphy and speech features are too complex to use without some dimension reduction. However, actigraphy and speech offer insight into patient experience both because they reflect a novel behavioral measure and because they involve limited self-assessment, which is known to be susceptible to psychological biases. Here we showed that we were able to quantify and select features from these objective measures in a preprocessing step, and then incorporate them into a clustering analysis. We found that one benefit of this approach is that these types of features indeed add dimensionality to a state solution, and the preprocessing in this case allowed for the derived features to add some biological interpretation. Additionally, we identified speech features that capture negative sentiment, possibly augmenting the ability for the states to detect disability versus wellness as indicated by higher correlation values between those states and the independent assessments. 

\subsection{Conclusions} 
Ultimately, this analysis combined AI and clinical knowledge to successfully reduce complex mobile data into useful health states that reflect important clinical time points and changes in patient experience (Figure 6). While all approaches should be tested and verified broadly across additional populations and data sets, this approach lays a solid foundation by which complex datastreams may be reduced into and authenticated as useful wellness information. We were able to show that we could successfully use this method in patients undergoing treatment for chronic pain, with results yielding new, distinct representations of patient experience. These findings imply it is possible to expand this approach to other illnesses associated with heterogeneous sets of symptoms. Finally, while we were able to compare our findings to known metrics, the health states provide deep insights in and of themselves that could aid a clinician in medical decision making and patient care. Given the growing use of digital health solutions, this approach to define Pain Patient States holds great promise in harnessing AI-driven solutions to aid in the care of large groups of chronic pain patients.

\section*{Acknowledgment}
The NAVITAS and ENVISION Studies Physician Author Group includes  Richard Rauck (The Center for Clinical Research), Eric Loudermilk (PCPMG Clinical Research Unit), Julio Paez (South Lake Pain Institute), Louis Bojrab (Forest Health Medical Center), John Noles (River Cities Interventional Pain), Todd Turley (Hope Research Institute), Mohab Ibrahim (Banner University Medical Center), Amol Patwardhan (Banner University Medical Center), James Scowcroft (KC Pain Centers), Rene Przkora (University of Florida), Nathan Miller (Coastal Pain and Spinal Diagnostics), and Gassan Chaiban (Ochsner Clinic Foundation).

The Boston Scientific Research Scientists Consortium includes Dat Huynh (Boston Scientific, Data Research and Engineering), Kristen Lechleiter (Clinical Research, Boston Scientific), Brad Hershey (Data Research and Engineering, Boston Scientific), Rex Woon (Data Research and Engineering, Boston Scientific), and Matt McDonald (Boston Scientific, Data Research and Engineering). 

We wish to acknowledge work by Erhan Bilal (IBM, Digital Health) for his work on consensus clustering. 


\bibliographystyle{IEEEtran}
\bibliography{IEEE_DH_refs}

\begin{thebibliography}{10}
\providecommand{\url}[1]{#1}
\csname url@samestyle\endcsname
\providecommand{\newblock}{\relax}
\providecommand{\bibinfo}[2]{#2}
\providecommand{\BIBentrySTDinterwordspacing}{\spaceskip=0pt\relax}
\providecommand{\BIBentryALTinterwordstretchfactor}{4}
\providecommand{\BIBentryALTinterwordspacing}{\spaceskip=\fontdimen2\font plus
\BIBentryALTinterwordstretchfactor\fontdimen3\font minus
  \fontdimen4\font\relax}
\providecommand{\BIBforeignlanguage}[2]{{%
\expandafter\ifx\csname l@#1\endcsname\relax
\typeout{** WARNING: IEEEtran.bst: No hyphenation pattern has been}%
\typeout{** loaded for the language `#1'. Using the pattern for}%
\typeout{** the default language instead.}%
\else
\language=\csname l@#1\endcsname
\fi
#2}}
\providecommand{\BIBdecl}{\relax}
\BIBdecl

\bibitem{Ekelund2009}
U.~Ekelund, S.~Brage, S.~J. Griffin, and N.~J. Wareham, ``{Objectively measured
  moderate- and vigorous-intensity physical activity but not sedentary time
  predicts insulin resistance in high-risk individuals},'' \emph{Diabetes
  Care}, vol.~32, no.~6, pp. 1081--1086, 2009.

\bibitem{Smirnova2020}
E.~Smirnova, A.~Leroux, Q.~Cao, L.~Tabacu, V.~Zipunnikov, C.~Crainiceanu, J.~K.
  Urbanek, and A.~Newman, ``{The Predictive Performance of Objective Measures
  of Physical Activity Derived from Accelerometry Data for 5-Year All-Cause
  Mortality in Older Adults: National Health and Nutritional Examination Survey
  2003-2006},'' \emph{Journals of Gerontology - Series A Biological Sciences
  and Medical Sciences}, vol.~75, no.~9, pp. 1779--1785, 2020.

\bibitem{Agurto2020}
\BIBentryALTinterwordspacing
C.~Agurto, G.~A. Cecchi, R.~Norel, R.~Ostrand, M.~Kirkpatrick, M.~J. Baggott,
  M.~C. Wardle, H.~de~Wit, and G.~Bedi, ``{Detection of acute
  3,4-methylenedioxymethamphetamine (MDMA) effects across protocols using
  automated natural language processing},'' \emph{Neuropsychopharmacology},
  vol.~45, no.~5, pp. 823--832, apr 2020. [Online]. Available:
  \url{https://doi.org/10.1038/s41386-020-0620-4}
\BIBentrySTDinterwordspacing

\bibitem{Corcoran2018a}
C.~M. Corcoran, F.~Carrillo, D.~Fern{\'{a}}ndez-Slezak, G.~Bedi, C.~Klim, D.~C.
  Javitt, C.~E. Bearden, and G.~A. Cecchi, ``{Prediction of psychosis across
  protocols and risk cohorts using automated language analysis},'' \emph{World
  Psychiatry}, vol.~17, no.~1, pp. 67--75, 2018.

\bibitem{Eyigoz2020}
\BIBentryALTinterwordspacing
E.~Eyigoz, S.~Mathur, M.~Santamaria, G.~Cecchi, and M.~Naylor, ``{Linguistic
  markers predict onset of Alzheimer's disease},'' \emph{EClinicalMedicine},
  vol.~28, p. 100583, nov 2020. [Online]. Available:
  \url{https://doi.org/10.1016/j.eclinm.2020.100583}
\BIBentrySTDinterwordspacing

\bibitem{Yu2019}
Y.~Yu, M.~Li, L.~Liu, Y.~Li, and J.~Wang, ``{Clinical big data and deep
  learning: Applications, challenges, and future outlooks},'' in \emph{Big Data
  Mining and Analytics}, 2019, pp. 288--305.

\bibitem{Ambigavathi2018}
M.~Ambigavathi and D.~Sridharan, ``{Big Data Analytics in Healthcare},'' in
  \emph{2018 10th International Conference on Advanced Computing, ICoAC 2018},
  2018.

\bibitem{Dinov2016}
I.~D. Dinov, ``{Methodological challenges and analytic opportunities for
  modeling and interpreting Big Healthcare Data},'' \emph{GigaScience}, vol.~5,
  no.~1, pp. s13\,742--016--0117--6, 2016.

\bibitem{Reinen2018}
J.~M. Reinen, O.~Y. Chen, R.~M. Hutchison, B.~T. Yeo, K.~M. Anderson, M.~R.
  Sabuncu, D.~Ongur, J.~L. Roffman, J.~W. Smoller, J.~T. Baker, and A.~J.
  Holmes, ``{The human cortex possesses a reconfigurable dynamic network
  architecture that is disrupted in psychosis},'' \emph{Nature Communications},
  vol.~9, no.~1, pp. 1--15, 2018.

\bibitem{Zelaya2019}
\BIBentryALTinterwordspacing
C.~E. Zelaya, J.~M. Dahlhamer, J.~W. Lucas, and E.~M. Connor, ``{Chronic Pain
  and High-impact Chronic Pain Among U.S. Adult},'' NCHS Data Brief 2020, Tech.
  Rep., 2019. [Online]. Available:
  \url{https://www.cdc.gov/nchs/products/index.htm.}
\BIBentrySTDinterwordspacing

\bibitem{Gatchel2007}
R.~J. Gatchel, Y.~B. Peng, M.~L. Peters, P.~N. Fuchs, and D.~C. Turk, ``{The
  Biopsychosocial Approach to Chronic Pain: Scientific Advances and Future
  Directions},'' \emph{Psychological Bulletin}, vol. 133, no.~4, pp. 581--624,
  2007.

\bibitem{Mullin2021}
S.~Mullin, J.~Zola, R.~Lee, J.~Hu, B.~MacKenzie, A.~Brickman, G.~Anaya,
  S.~Sinha, A.~Li, and P.~L. Elkin, ``{Longitudinal K-means approaches to
  clustering and analyzing EHR opioid use trajectories for clinical
  subtypes},'' \emph{Journal of Biomedical Informatics}, vol. 122, p. 103889,
  2021.

\bibitem{Lotsch2018}
J.~L{\"{o}}tsch and A.~Ultsch, ``{Machine learning in pain research},''
  \emph{Pain}, vol. 159, no.~4, pp. 623--630, 2018.

\bibitem{Jenssen2021}
M.~D.~K. Jenssen, P.~A. Bakkevoll, P.~D. Ngo, A.~Budrionis, A.~J. Fagerlund,
  M.~Tayefi, J.~G. Bellika, and F.~Godtliebsen, ``{Machine learning in chronic
  pain research: A scoping review},'' \emph{Applied Sciences (Switzerland)},
  vol.~11, no.~7, p. 3205, 2021.

\bibitem{Reinen2020}
J.~Reinen, S.~Berger, C.~Agurto, R.~Ostrand, E.~Loudermilk, J.~Paez, G.~Cecchi,
  J.~Rogers, K.~Lechleiter, and R.~Rauch, ``{Defining Multi-Dimensional Dynamic
  States of Chronic Pain Using a Mobile Clinical Platform},'' in \emph{World
  Institute of Pain}, 2020.

\bibitem{Anitescu2022}
M.~Anitescu, A.~Antony, R.~Rauck, E.~Loudermilk, J.~Paez, L.~Bojrab, J.~Noles,
  T.~Turley, M.~Ibrahim, A.~Patwardhan, J.~Scowcroft, R.~Przkora, N.~Miller,
  G.~Chaiban, D.~Huynh, K.~Lechleiter, B.~Hershey, R.~Woon, J.~Reinen,
  C.~Agurto, G.~Cecchi, J.~Rogers, and M.~McDonald, ``{Patient States:
  Artificial Intelligence-Driven Metric Providing Comprehensive Yet
  Straightforward Understanding of Chronic Pain Patients.}'' in \emph{North
  American Neuromodulation Society}, 2022.

\bibitem{Group1990}
T.~E. Group, ``{EuroQol - a new facility for the measurement of health-related
  quality of life},'' \emph{Health policy}, vol.~16, no.~3, pp. 199--208, dec
  1990.

\bibitem{Fairbank2000}
\BIBentryALTinterwordspacing
J.~C.~T. Fairbank and P.~B. Pynsent, ``{The Oswestry Disability Index},''
  \emph{Spine}, vol.~25, no.~22, pp. 2940--2953, nov 2000. [Online]. Available:
  \url{http://journals.lww.com/00007632-200011150-00017}
\BIBentrySTDinterwordspacing

\bibitem{Pennebaker2015}
J.~W. Pennebaker, R.~L. Boyd, K.~Jordan, and K.~Blackburn, ``{The development
  and psychometric properties of LIWC2015},'' \emph{Austin, TX: University of
  Texas at Austin}, 2015.

\bibitem{Eyben2010}
\BIBentryALTinterwordspacing
F.~Eyben, M.~W{\"{o}}llmer, and B.~Schuller, ``{OpenSMILE - The Munich
  versatile and fast open-source audio feature extractor},'' in \emph{MM'10 -
  Proceedings of the ACM Multimedia 2010 International Conference}.\hskip 1em
  plus 0.5em minus 0.4em\relax New York, New York, USA: ACM Press, 2010, pp.
  1459--1462. [Online]. Available:
  \url{http://dl.acm.org/citation.cfm?doid=1873951.1874246}
\BIBentrySTDinterwordspacing

\bibitem{DeJong2009}
N.~H. de~Jong and T.~Wempe, ``{Praat script to detect syllable nuclei and
  measure speech rate automatically},'' \emph{Behavior Research Methods},
  vol.~41, no.~2, pp. 385--390, 2009.

\end{thebibliography}

\end{document}